\documentclass[sigconf,natbib=true,anonymous=false
]{acmart}

\AtBeginDocument{%
  \providecommand\BibTeX{{%
    \normalfont B\kern-0.5em{\scshape i\kern-0.25em b}\kern-0.8em\TeX}}}

\renewcommand\footnotetextcopyrightpermission[1]{} %

\copyrightyear{2023} 
\acmYear{2023} 
\acmPrice{15.00}
\acmDOI{10.1145/3477495.3531769}
\acmISBN{978-1-4503-8732-3/22/07}

\usepackage{graphicx}
\usepackage{enumitem}
\usepackage{lipsum}
\usepackage{float}
\usepackage{subfigure}
\usepackage{multirow,makecell}
\usepackage{pifont}
\usepackage{caption}
\usepackage{subcaption}
\usepackage{balance}
\usepackage{adjustbox}
\usepackage{dsfont} 
\usepackage{arydshln} %
\usepackage{soul}
\usepackage{amsmath} %
\usepackage{stfloats}
\usepackage{booktabs}

\newcommand{\cmark}{\ding{51}}%
\newcommand{\xmark}{\ding{55}}%

\newcommand{\AK}[1]{\textcolor{blue}{{\bf [AK: }{\bf ]}}}
\newcommand{\EK}[1]{\textcolor{green}{{\bf [EK: }{\bf ]}}}
\newcommand{\AY}[1]{\textcolor{pink}{{\bf [AY: }{\bf ]}}}

\begin{document}
\fancyhead{}

\title{Constructing Set-Compositional and Negated Representations \\ for First-Stage Ranking}

\author{Antonios Minas Krasakis}
\email{a.m.krasakis@uva.nl}
\affiliation{%
  \institution{University of Amsterdam}
  \country{The Netherlands}
}
\author{Andrew Yates}
\email{a.c.yates@uva.nl}
\affiliation{%
  \institution{University of Amsterdam}
  \country{The Netherlands}
}
\author{Evangelos Kanoulas}
\email{e.kanoulas@uva.nl}
\affiliation{%
  \institution{University of Amsterdam}
  \country{The Netherlands}
}

\renewcommand{\shortauthors}{Krasakis, et al.}

\begin{abstract}

Set compositional and negated queries are crucial for expressing complex information needs and enable the discovery of niche items like ``Books about non-European monarchs''. %
Despite the recent advances in LLMs, first-stage ranking remains challenging due to the requirement of encoding documents and queries independently from each other. This limitation calls for constructing compositional query representations that encapsulate logical operations or negations,
and can be used to
match relevant documents effectively. 
In the first part of this work, we explore constructing such representations in a zero-shot setting using vector operations between lexically grounded Learned Sparse Retrieval (LSR) representations.
Specifically, we introduce \textit{Disentangled Negation} that penalizes only the negated parts of a query, and a \textit{Combined Pseudo-Term} approach that enhances LSR’s ability to handle intersections.
We find that our zero-shot approach is competitive and often outperforms retrievers fine-tuned on compositional data, highlighting certain limitations of LSR and Dense Retrievers.
Finally, we address some of these limitations and improve LSR's representation power for negation, by allowing them to {attribute negative term scores} and effectively penalize documents containing the negated terms.

\end{abstract}

\maketitle

\section{Introduction}

Information retrieval (IR) systems have become increasingly important in the era of Large Language Models (LLMs)~\cite{asai2023retrieval,lewis2020retrieval,izacard2020leveraging}. Search plays a crucial role in helping LLMs answer user questions or engage in information-seeking conversations by providing evidence to ground or verify their responses, thereby reducing the risk of hallucinations.
As these technologies and applications evolve, search queries become increasingly complex~\cite{yang2018hotpotqa}. One common type of complexity arises when a query contains multiple constraints, often combined using logical operations like conjunctions (``and''), disjunctions (``or''), or negations (``not''). For example, a user might search for books about French monarchs but exclude Napoleon, or look for a documentary about Canada filmed in Mississippi. These complex queries help users discover niche items, and can support tasks like known-item retrieval~\cite{arguello2021tip,hagen2015corpus,bhargav2022s} or conversational search and clarifying questions~\cite{krasakis2020analysing}.

Traditionally, IR systems used boolean queries with logical operations~\cite{frants1999boolean}, but those were superseded by probabilistic and vector space models without any Boolean logic.
Modern SOTA Neural IR retrievers also lack Boolean logic since they are pre-trained on large-scale ad-hoc training datasets such as MSMarco~\cite{nguyen2016ms} that contain few of these logical operators and negations,
and little work has been done to evaluate their effectiveness there.

Recent research found that even state-of-the-art retrievers struggle with conjunctions and negations ~\cite{malaviya2023quest,weller2023nevir}. 
For negations in particular, previous work highlighted that one of the major challenges there concerns that the positive and negative parts of the query often interfere with each other, causing negation methods to harm the ranking scores of relevant documents~\cite{widdows2003orthogonal,dunlop1997effect}.
First stage ranking is particularly challenging, due to the requirement of encoding queries and documents independently (ie.~no cross-attention) in a vector representation before matching them using a similarity metric.
This limitation necessitates that vector representations must effectively encapsulate this compositionality.
In the space of Dense Retrieval (DR), most work has focused on negation, where \citet{widdows2003orthogonal} introduced the method of Orthogonal Negation, targeted at removing only the negated aspects of a dense vector.
For lexical retrieval, most methods focused on applying Rocchio relevance feedback~\cite{rocchio1971relevance} to "dampen" the appearance of negated terms in retrieved documents~\cite{salton1990improving,zhang2013query}.

In this work we focus on lexical representations of queries and documents. Lexical methods are  interpretable, since they build representations that are grounded to a vocabulary. %
This provides the necessary flexibility in allowing us to design intuitive operations on the representations that can encapsulate set-compositionality.
Meanwhile, recent research has overcome most limitations of traditional lexical methods, ie. the vocabulary mismatch and the ability to do supervised learning~\cite{formal2021splade,nguyen2023unified,zhuang2021tilde,lin2021few,mallia2021learning}, with Learned Sparse Retrievers (LSR) %
estimating term importance in the context of a specific query or document.
As a consequence, LSR models are very effective first-stage rankers (SOTA in out-of-domain settings and competitive to DR in-domain), while preserving this lexical grounding in their representations.

We first propose a zero-shot framework that constructs set-compositional representations of queries by performing Linear Algebra Operations (LAO) between LSR representations. 
We design LAO that manipulate sparse representations to reflect logical operations. %
For negations, we propose the method of \textit{Disentangled Negation}, that takes advantage of this lexical grounding and the "disentanglement" between different dimensions of the representation to mitigate the pressing problem of interference between positive and negative aspects of a query~\cite{dunlop1997effect,widdows2003orthogonal}. Regarding intersections, we identify the shortcoming of current LSR representations in capturing joint term occurrences and conclude this has a positive impact in ranking.
Such zero-shot methods are useful as they can be used with existing pre-trained models without the need of further finetuning, giving users more fine-grained control over the ranking, or assist in collecting training data (ie. top-K pooling~\cite{voorhees2005trec}) to train retrievers for compositionality.

Beyond zero-shot methods, we investigate the capabilities of retrievers in learning such compositional representations from training on the ranking task. We find that both LSR and Dense retrievers are effective on negations but not intersections.%
We find that performance of the supervised retrievers are comparable to the zero-shot LAO, and we observe that Disentangled Negation appears to have a slight advantage in top ranking positions. Our hypothesis is that this is due to a limitation of the (supervised) LSR models, which is that they can remove negated terms from queries but they cannot penalize documents that contain them. 
To address this limitation, we adapt the architecture of LSR models to allow them attribute negative term scores. 
Our experiments demonstrate that in settings where enough training data exists, this is very effective and leads to large improvements compared to traditional LSR models that do not produce negative weights.
For negations, we observe an advantage of methods that have the capacity to score negatively (ie. penalize) terms.
Intrigued by this observation, we explore to what extent different compositionality types can be learned from training data for different retrieval families. 
We conclude that both Dense (DR) and Learned Sparse Retrievers (LSR) do benefit from training on negations, but somewhat surprisingly  cannot learn how to handle Intersections (at least under domain shift). %

In particular, we make the following contributions:
\begin{itemize}
    \item We propose a zero-shot framework that constructs set - compositional representations of queries by performing Linear Algebra Operations (LAO) between atomic representations. Most importantly, we introduce the LAO of \textit{Disentangled Negation}, that penalizes negated terms of a query without affecting the positive ones, and Combined Pseudo-Terms, that expands the representation capabilities of LSR models when modelling Intersections.
    \item We explore the ability of Dense and Learned Sparse Retrievers to learn compositional representations, concluding that this is effective for Set Difference, but none of the models is able to do so for Intersections.
    \item We improve the representation power of Learned Sparse Retrievers (LSR) for negated queries, by allowing them to attribute negative term scores and penalize documents containing the negated terms.
\end{itemize}

To the best of our knowledge, this is the first work that explores constructing set-compositional and negated representations using LSR models that are lexically grounded, SOTA first-stage rankers.

\section{Related Work}\label{sec:RW}

We first discuss the use of logical operators in traditional IR systems, followed by related work on compositionality and transformers, and close this section with works discussing negations for IR.

\subsubsection*{Logical operations in IR}
Early Information Retrieval systems used boolean search, employing logical operators to construct queries and retrieve documents based on term disjunction, conjunction, or absence~\cite{salton1972dynamic,mcgill1976syracuse,rocchio1971smart}. However, these systems didn't rank results, returning only a document set, and required a complex query language to express queries~\cite{frants1999boolean, rocchio1971smart, salton1972dynamic},
ultimately declining in popularity as IR shifted to statistical methods like tf-idf and BM25~\cite{hiemstra2000probabilistic,robertson2009probabilistic}.

Later approaches attempted to incorporate compositionality and logic operations by performing linear algebra on dense vector representations~\cite{widdows2021should,widdows2003orthogonal} or applying relevance feedback to lexical methods~\cite{salton1990improving, basile2011negation,zhang2013query, widdows2003word}.
Among the former, \citet{widdows2003orthogonal} introduced orthogonal negation, which subtracts the projection of a negative vector from the positive vector. %
Experiments showed that this method is effective in removing negated concepts, but may still negatively impact retrieval of the positive ones.
\citet{widdows2021should} surveyed various vector composition methods, distinguishing between explicit approaches (e.g., orthogonal negation) and implicit methods (e.g., attention in neural networks).
Regarding relevance feedback methods, \citet{salton1990improving} used Rocchio feedback~\cite{rocchio1971relevance} to penalize terms found in non-relevant document sets and handle negations. 
\citet{zhang2013query} proposed a similar method for session search, introducing an algorithm that models a session using a theme query vector $\vec{q}$, along with positive ($+\Delta \vec{q}$) and negative ($-\Delta \vec{q}$) query changes based on user document clicks and query modifications.
To the best of our knowledge, very few studies  investigate compositionality and negations for modern neural IR models. 
We draw inspiration from these works and explore how some of their ideas can be applied to LSR, that are state-of-the-art first-stage rankers%
.
\subsubsection*{Compositionality and negation in Transformers}
Modern Neural IR models are based on the transformer architecture~\cite{vaswani2017attention,karpukhin2020dense,khattab2020colbert,formal2021splade,izacard2021unsupervised}. However, transformers' ability to tackle compositionality and negations has been widely debated in the NLP community, with many works highlighting their struggle in doing so
~\cite{baan2019realization,gubelmann2022context,kassner2019negated,hossain2020analysis,peng2024limitations,dziri2024faith}. 
Our work differs in two key axes. First, we explore set-compositionality specifically for ranking, rather than tasks involving numerical operations or problem decomposition. Second, we focus on first-stage ranking, where documents and queries must be encoded independently. This constraint requires models to encapsulate compositionality within a single vector, without relying on multiple layers of cross-attention.
Word2Box~\cite{dasgupta2021word2box} introduced n-dimensional hyperrectangle (box) embeddings, showing that region-based approaches outperform traditional embeddings for set-compositionality and polysemy in word similarity benchmarks. However, these would require integration over boxes to compute query and document scores, which makes them difficult to apply to first-stage ranking that relies on kNN search~\cite{douze2024faiss,yates2021pretrained}. 
Therefore, we focus on the first-stage setting and investigate the widely used DR and LSR models in the context of Neural IR.
\subsubsection*{Challenges for negations in IR}
Previous work discusses the effectiveness of negation in IR and highlights the challenges therein. 
Previous works showed that applying negations can reduce method effectiveness, as negating a concept closely linked to the target information need significantly impacts the ranking of the relevant items~\cite{widdows2003orthogonal,dunlop1997effect}. 
To address this, a set of methods try to limit the impact of negated terms or aspects in a query and control it using a hyper-parameter (Constant subtraction)~
\cite{kowalski2007information,koopman2014understanding,widdows2003orthogonal}. %
Research in conversational search and clinical IR finds evidence conflicting with our intuitions regarding how to handle negation, concluding that including negated terms can have positive effects in retrieval~\cite{koopman2014understanding,krasakis2020analysing}.
Lastly, \citet{mcquire1998ambiguity} discusses the problem of ambiguity in queries with negation. They show that when such queries increase in complexity, disagreements arise on attributing the negation between human annotators, and automatic systems failed as well. 
In this work, we build on these findings and further investigate negations in the context of Neural IR and LSR models.

\section{Methodology}

In this section we outline our methodology. We start with preliminaries on LSR (Section \ref{sec:method:lsr}), %
followed by a framework for deriving set-compositional representations (Section \ref{sec:method:rq1}). 
On Section \ref{sec:method-rq3} we introduce an adaptation of LSR for modelling negations.

\subsection{Learned Sparse Retrievers}\label{sec:method:lsr}
Learned Sparse Retrieval (LSR) is a family of efficient and effective Neural IR models.
They improve upon traditional lexical methods (eg. BM25~\cite{robertson2009probabilistic}) 
by (a) learning an expansion over the input text, closing the vocabulary mismatch, 
and (b) predicting term importance scores. 
Splade~\cite{formal2021splade} does this by computing term importance scores $w_j$ for a token $t_j$ at position $j$ over a WordPiece vocabulary.
To get $w_j$, Splade passes input embeddings through multiple transformer layers, deriving an token representation $h_i$ at input position $i$. 
Then it uses a MLM head ($E$, $b$) to compute a weight $out_ij$ corresponding to the importance of vocabulary term $j$ given input term $i$. 
\begin{equation}
    out_{ij} = transform(h_i)^T \cdot E_j + b_j \quad j \in \{1,...,|V|\} 
\label{eqn:splade-output-vanilla}
\end{equation}
A Rectified Linear Unit and a logarithmic function ensure term scores are positive with a log-saturation effect. Finally, scores are aggregated over all input tokens with a Max Pooling operation:
\begin{equation}
w_j= \max_{i \in t} \log \left(1 + \text{ReLU}(out_{ij}) \right)
\label{eqn:splade-activation-vanilla}
\end{equation}
In practice, this happens in a query or document encoder $f_{QE}, f_{DE}$ that maps the input text to a sparse representation $\vec{Q}$ or $\vec{D}$.
\begin{align}
\begin{split}
\vec{Q} = f_{QE}(Q) = [w_0^q, w_1^q, ..., w_{|V|}^q ] \\
\vec{D} = f_{DE}(D) = [w_0^d, w_1^d, ..., w_{|V|}^d ] 
\end{split}
\end{align}
For brevity, these representations can also be written as 
$\vec{Q} = (t_j: w_j, ...) \quad \forall w_j \neq 0$ .
The ranking score between a query and document is computed by multiplying the term scores between $Q$ and $D$ across vocabulary $V$.
\begin{equation}\label{eq:splade-scoring}
score(Q,D) = 
\sum_{i=1}^{|V|} 
w_i^q \cdot w_i^d
\end{equation}

\subsection{Deriving compositional query representations with Linear Algebra Operations (LAO)}\label{sec:method:rq1}

\begin{figure*}[]
    \centering
    \includegraphics[width=0.9\textwidth]{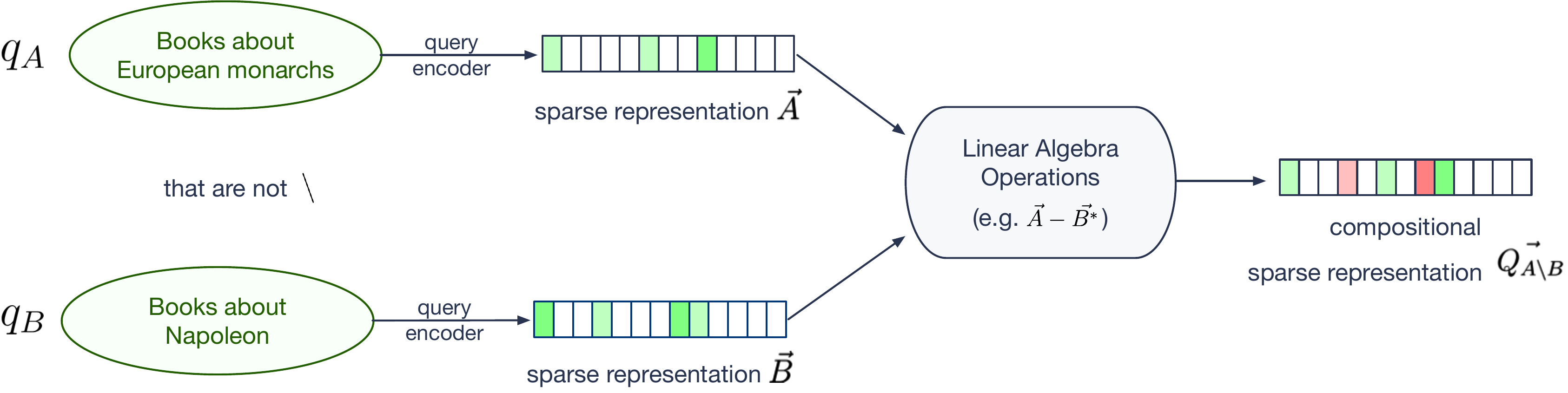}
    \caption{A zero-shot framework for constructing compositional query representations with Linear Algebra Operations.}
    \label{fig:framework-lao}
\end{figure*}

In this part we describe a framework (Figure \ref{fig:framework-lao}) to derive representations of set-compositional queries, based on their atomic queries and set operations (Section \ref{sec:method:rq1:framework}). The framework relies on Linear Algebra Operations (LAO) that are designed to tackle a particular logical operations (Section \ref{sec:method:rq1:lao}).

\textit{Notation.}
Let us consider an atomic query $q_A$, with a sparse query representation vector $\vec{A} = [w_0^A, w_1^A, ..., w_{|V|}^A] 
$
and let $|A|$ be the set of relevant documents to $q_A$. 
Let $q_{A \star B}$ be a compositional query, where $\star \in \{ \setminus , \cup, \cap, \}$ for which we seek to construct a representation $\vec{Q_{A \star B}}$ that is effective in matching the relevant documents $|A \star B|$.
For instance, consider the atomic queries \textsc{$q_A$=``Birds of Colombia''} and \textsc{$q_B$=``Birds of Venezuela''}
with the binary query representation vectors 
$\vec{A}=$ \textsc{(birds:$1$, fly:$1$, Colombia:$1$, Andes:$1$)} and $\vec{B}= $\textsc{(birds:$1$, fly:$1$, Venezuela:$1$, Andes:$1$)}. 
\subsubsection{A framework for deriving zero-shot compositional representations}\label{sec:method:rq1:framework}
Current query encoders have been pre-trained on large ad-hoc query datasets that contain few set-compositional queries with \textit{union}, \textit{intersection} and \textit{set difference}. 
Therefore, our hypothesis is that they can construct effective representations for atomic queries, which we can later combine using LAO designed to reflect the logical set-operations (further motivated in Section \ref{sec:method:rq1:lao}).

Our framework requires an explicit formulation of set-compositional queries, ie. the atomic queries and logical operations. 
For the set-compositional query $q_{A\cap B} = $ \textsc{(Birds of Colombia)} \textsc{that are also} \textsc{(Birds of Venezuela)}, a decomposition is required to derive atomic queries \textsc{$A$=(Birds of Colombia)}, \textsc{$B$=(Birds of Venezuela)}, and the set-compositional operation  of \textit{Intersection} ($A \cap B$). 
We do not investigate how to derive this decomposition here, as preliminary experiments showed that in this dataset, LLMs can perform this task very accurately. Instead, we use the atomic queries and operations provided directly by the QUEST dataset. 
Then, we use an LSR query encoder $f_{QE}$ to
construct the atomic query representations $\vec{A}$, $\vec{B}$ from the corresponding atomic queries $q_A$, $q_B$:
\[\vec{A} = f_{QE} (q_A)\]
Then, we use a compositional query encoder $f_{CQE}$ to combine the atomic embeddings using linear algebra operations that reflect the logical set operations. 
\[ f_{CQE}(q_A \star q_B) = \vec{A} \circ \vec{B} 
\quad 
\text{, where }
\star \in \{ \cap, \cup, \setminus \}
\]
and $\circ$ is the corresponding LAO described in Subsection \ref{sec:method:rq1:lao}.

\subsubsection{Linear Algebra Operations (LAO) for set-theoretic operations}\label{sec:method:rq1:lao}
In this part, we define LAO between sparse embeddings, that reflect the set operations of \textit{union}, \textit{intersection} and \textit{difference}. 

\underline{\textbf{Set Difference ($q_A \setminus q_B$)}.}
Set difference is closely related with negation, since it includes a positive and a negative part within a query. 
Based on set theory, we are looking for all elements (documents) of $A$ that are not members of $B$, ie.
    $|A \setminus B| = |A| - |A \cap B|$
.
An intuitive formulation to representing set difference would be \textit{Subtraction}, penalizing documents containing terms related to $q_B$ and therefore $\vec{B}$:

\[ 
f_{CQE}(A \setminus B) = \vec{A} - \vec{B}\]
Considering the previous example, simple Subtraction would result in a representation of \textsc{(Colombia:$1$, Venezuela:\textbf{-}1)}, penalizing documents mentioning \textsc{Venezuela}, but also
removing \textsc{birds} and \textsc{fly}.
Prior work highlighted the difficulty of building negated representations using subtraction, due to an interference between positive and negative query
~\cite{widdows2003orthogonal,dunlop1997effect}.
Intuitively, \citet{widdows2003orthogonal} explains that when queries $q_A$,$q_B$ have high semantic overlap, negating $\vec{B}$ inevitably pushes $A$ away from it's original representation. 
LSR representations present an opportunity to tackle this problem, since they are sparse and lexically grounded, 
as each dimension $i$ preserves a semantic relationship to token $t^i$.
Therefore LSR representations are much more disentangled compared to DR, allowing us to formulate a representation that only penalizes dimensions present in $B$ without affecting dimensions relevant to $A$, expressing \textit{Disentangled Negation}:
\begin{equation}
    f_{CQE} (A \setminus B) = \vec{A} - \vec{B^*} 
\quad\text{,where } 
\vec{B^*} = \vec{B} - \mathds{1}_A \odot \vec{B}
\end{equation}
The final representation \textsc{(Colombia:$1$,  Venezuela:$-1$, birds:$1$, fly:$1$, Andes:$1$)}, manages to penalize terms related to negative query aspects while retaining all of the positive ones.

\underline{\textbf{Intersection ($q_A \cap q_B$)}.}
Expressing query term intersections in LSR can be challenging, due to the semantic relationship of each dimension to a token.
Our hypothesis is that for modeling intersections, it would be useful to 
expand the representation vector from $|V|$ to $|V|^2$ and create \textbf{Combined Pseudo-Terms (CPT)}, capturing joint term-occurence such as \textsc{Colombia}$\cap$\textsc{Venezuela}.
In practice, this is an outer product between query vectors $\vec{A}$ and $\vec{B}$:
\begin{equation}\label{eq:intersection-outer-product}
f_{CQE}(A \cap B) = \vec{A} \otimes \vec{B} = 
\begin{bmatrix}
w_{{A^0}{B^0}} & w_{{A^0}{B^1}} & ... & w_{{A^0}{B^{|V|}}} \\
w_{{A^1}{B^0}} & w_{{A^1}{B^1}} & ... & w_{{A^1}{B^{|V|}}} \\
... & ... & ... & ... \\
w_{A^{{|V|}}{B^0}} & ... & ... & w_{{A^{|V|}}{B^{|V|}}}
\end{bmatrix}
\end{equation}
where $w_{{A^i}{B^j}} = \sqrt{w_{A}^{i} \cdot w_{B}^{j}}$. For efficiency reasons, we limit the number of dimensions (top-terms per compositional query) to $5$.

Similarly, we represent documents with the outer product of the document vector: 
\begin{equation}\label{eq:intersection-outer-product-document}
f_{CDE} (D) = \vec{D} \otimes \vec{D} 
\quad \text{,where } w_{{D^i}{D^j}} = \sqrt{w_{D}^{i} \cdot w_{D}^{j}}
\end{equation}

Finally, we compute the \textit{Combined Pseudo-Scores} of queries and documents as:
\begin{equation}\label{eq:intersection-outer-product-score}
CPTscore(A \cap B, D) = || f_{CQE}(A \cap B) \cdot f_{CDE}(D) ||
\end{equation}

The quadratic expansion of dimensions %
expands the index size, but can be supported by modern inverted indexes designed for LSR models~\cite{bruch2024efficient}.
Alternatively, a two-step approximation selecting a collection subset (similar to ~\cite{khattab2020colbert}) and computing the expanded representation ($\vec{D} \otimes \vec{D}$) on the fly would also be possible.

\underline{\textbf{Union ($q_A \cup q_B$)}.}
Based on the Inclusion-Exclusion principle~\cite{pinter1976set} of set theory, ie.
$|A\cup B| = |A| + |B| - |A \cap B|$
, an approximate representation of union between embeddings can be formulated as \textit{Addition}: 
\begin{equation}\label{eq:union-addition}
f_{CQE}(A \cup B) = \vec{A} + \vec{B}
\end{equation}

However, set theory is concerned with boolean set membership, while for ranking term weights  $w_A^i$ are important for calculating relevance scores
. 
Therefore, such an approximation would tend to inflate scores of common terms between $A$ and $B$%
, as those would be counted twice. To avoid this, we also consider a \textit{MaxPooling} operation for union between atomic embeddings:
\begin{equation}\label{eq:union-maxpool}
f_{CQE}(A \cup B) = MaxPool(\vec{A} , \vec{B})
\end{equation}

\subsection{Improving LSR representations for negations}\label{sec:method-rq3}

Traditionally, lexical and LSR models use positive term weights ($w_i$) to compute ranking scores between queries and documents (Eq. \ref{eq:splade-scoring}). We hypothesize that this limits their ability to handle negations, as they can't represent and penalize negated terms in documents.

\begin{figure}[]
    \centering
        \includegraphics[width=\columnwidth]{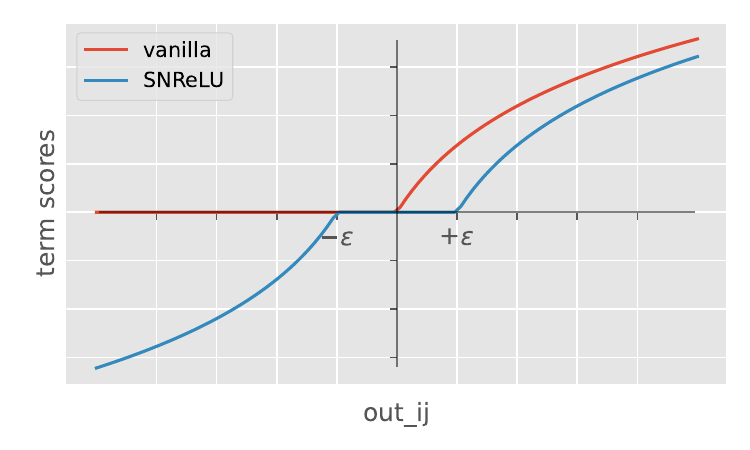}%
    \caption{Splade activation functions.}
    \label{fig:activation-functions}
\end{figure}

To address this, we modify the architecture of Splade, a SOTA LSR model, to allow \textbf{negative} term weights for representing negated aspects in queries or documents. We design a new activation function with properties similar to Splade's (Eq. \ref{eqn:splade-activation-vanilla}), such as the log saturation effect and sparsity, while also supporting negative term weights. Specifically, we introduce \text{SNReLU}, %
a symmetrical-odd version of Splade's activation function, which enables term sparsity ($0$ term scores) by shifting the original function by $-\epsilon$ and $+\epsilon$.

\begin{align}
\begin{split}
out\_pos_{ij} = \log (1 + \text{ReLU}(out_{ij}-\epsilon))\\
out\_neg_{ij} = -\log (1 + \text{ReLU}(-out_{ij} + \epsilon) )
\end{split}
\end{align}

We propose to aggregate token scores over the input in two different ways.
The first selects the input position with the largest absolute value ($\pm |max|$, Eq. \ref{eq:SNRelu-maxV2}).
Alternatively, we first pool the maximum positive and minimum negative values for each token, and then summing the two ($sum$ Eq. \ref{eq:SNRelu-maxV3}).

\begin{equation}\label{eq:SNRelu-maxV2}
w_{j}= 
\begin{cases}
    \max\limits_{i \in t} (out\_pos_{ij}) & , if
    \max\limits_{i \in t}|out\_pos_{ij}| \geq \max\limits_{i \in t}|out\_neg_{ij}| \\
    
    \min\limits_{i \in t} (out\_min_{ij}) & 
     ,otherwise
\end{cases}
\end{equation}

\begin{equation}\label{eq:SNRelu-maxV3}
w_{j} = \max\limits_{i \in t} (out\_pos_{ij}) + \min\limits_{i \in t} (out\_min_{ij})
\end{equation}

\section{Experimental Setup}
In this section we describe our experimental setup.

\subsection{Datasets and Evaluation}

\subsubsection*{QUEST}\label{sec:exper:datasets:quest}
QUEST~\cite{malaviya2023quest} is an entity retrieval dataset containing set-compositional queries (eg. ``Ming dynasty or Chinese historical novels'').
The dataset is build from multiple atomic queries, ie. individual atomic constraints that are combined using \textit{set difference}, \textit{union}, or \textit{intersection}.
Each query contains a template version that verbalizes the set operation using templates (eg. ``X that are also Y'' for intersection)
and a human-rephrased version.
Entity categories (eg. ``Ming dynasty novels'') are used as atomic queries and composed into a compositional template-based query, using templates. Human annotators paraphrase the template-based queries, while also assessing their fluency and naturalness. 
Throughout this work, we use the template versions of the queries, as during preliminary experiments we found no significant difference between the two.
We evaluate performance using $NDCG@10$, that evaluates retrieval effectiveness while giving more importance to the top ranking positions, and $R@100$, that evaluates what percentage of relevant documents is retrieved within the top-100 documents. 

\textit{Train-test split across domains.}
During initial experiments, we identified a leakage between the original train and test splits. 
Although compositional queries were unique across sets, their atomic queries appeared in both, and even when atomic queries were unique, they often had substantial overlap with other atomic queries. 
For example, train query ``Books about monarchs but not about France'' is very closely related to the test query ``Books about monarchs that don't include Napoleon'',
creating an overlap of $75\%$ between the two relevance sets.
In such cases, it is not clear whether the retriever learns to identify relevance and construct set compositional representations, or simply memorizes associations between atomic queries and relevant documents.
To solve this issue %
we perform a domain split ensuring each entity category appears either in train or test set, keeping all ``Film and ``Book'' category queries in the training set and ``Plants'' and ``Animals'' in the test set. 
\subsubsection*{NevIR}
NevIR~\cite{weller2023nevir} is a benchmark aiming to evaluate how negation impacts Neural Information Retrieval. 
It contains synthetic counterfactual query and document pairs, build on top of minimally contrasting documents. %
Contrasting documents are identical with only a clause being different, for instance: \textsc{``... Generally, drug possession is an arrestable offense, ...''} vs. \textsc{``..However, rarely is drug possession an arrestable offense...''}. Then, two queries are constructed to be relevant to one of the two documents.
Following the evaluation setup of \citet{weller2023nevir}, we 
use \textit{pair-wise accuracy} for evaluation, that measures whether both counterfactual queries scored the relevant document higher than the irrelevant.

\subsection{Retrieval models and Baselines}

First we describe the retrieval models used in our experiments, followed by baseline and ablation methods.

\subsubsection{Retrieval Models}
We include representative models from different families of retrieval models, specifically:
\begin{itemize}
    \item BM25~\cite{robertson2009probabilistic}, a SOTA lexical retriever (using pyserini~\cite{lin2021pyserini})
    \item Splade~\cite{formal2021splade,formal2021spladev2,lassance2024splade}, a SOTA Learned Sparse Retriever
    \item Contriever~\cite{izacard2021unsupervised}, a SOTA Dense Retriever
\end{itemize}

We start from the splade-v2\footnote{https://huggingface.co/naver/splade-cocondenser-ensembledistil} and contriever\footnote{https://huggingface.co/facebook/contriever-msmarco} checkpoints, unless stated otherwise. 
We perform a domain pre-finetuning step on the non-compositional subset of QUEST (questNC), to adapt models from the distribution of MSMarco to QUEST (eg. Wikipedia documents instead of marco passage collection). Note that during this step we only use atomic queries and not compositional queries.

\subsubsection{Compositionality-aware baselines and ablations}

We use appropriate baselines and linear algebra operation ablations designed for each logical set-operation.

\subsubsection*{Baselines}
    \begin{itemize}
    \item \textit{vanilla}: Verbalizes the set operation using a template per set operation (eg. ``$q_A$ that are not $q_B$'')
    \item \textit{NRF} ~\cite{salton1990improving}: Negative relevance feedback is based on the Rocchio feedback algorithm~\cite{rocchio1966document} and uses positive and negative feedback to change the query vector away from irrelevant documents and closer to relevant. 
    We adapt the method, using the negative query vector $\vec{B}$ as the negative feedback vector, ie. $\vec{A}-\lambda \cdot \vec{B}$.

    \item \textit{Orthogonal negation}~\cite{widdows2003orthogonal}: a method that subtracts the projection $\pi$ of query B from query A, which in practice alters the positive vector A to remove the common aspects found in B,
        ie. $A \setminus B = 
        \vec{A} - \pi_{\vec{B}} \vec{A} = 
        \vec{A} - \frac{\vec{A}\cdot\vec{B}}{|\vec{B}|^2}\cdot \vec{B}$%

    \item \textit{score-fusion}: we rank documents for queries $A$ and $B$ independently, and fuse the query-document scores using the corresponding operand $\circ$:
    \[score(d,q_A,q_B) = score(d,q_A) \circ score(d,q_B)\]
    where $\circ$ corresponds to $+$ for union, $\cdot$ for intersection and $-$ for negation. This is inspired by probability theory and based on the assumption that retrieval scores try to approximate the probability of relevance (eg. $P(A \cap B) = P(A) \cdot P(B)$)
    
    \item \textit{score-fusion-scaled}: similar to the previous, but with document scores scaled per atomic query.
    
    \end{itemize}
    
Among other possible baselines, \citet{zhang2013query} uses document clicks and query changes, which are unavailable in our setting. Although an adaptation to query vectors would be possible, it can be viewed as a special case of \textit{NRF}.

\subsubsection*{LAO ablations:}
\begin{itemize}
    \item \textit{Addition:} $\vec{A}+\vec{B}$, used for intersection and union (as motivated on Section \ref{sec:method:rq1:lao})
    
    \item \textit{Max Pool:} $MaxPool(\vec{A}+\vec{B})$, used for intersection and union (as motivated on Section \ref{sec:method:rq1:lao})

    \item \textit{Ignore Negation:} $\vec{A}$, used for set difference. Simply ignores the negated part of a query.
\end{itemize}

\section{Results}

In this Section, we discuss our experimental results and seek answers to the following Research Questions:

\begin{enumerate}[label=\textbf{RQ\arabic*},leftmargin=*]
    \item\label{rq:query-representation} 
    How can we build effective sparse representations of set-compositional queries in a zero-shot setting?
    
    \item\label{rq:learn-representations} To what extent can the different families of Dense and Learned Sparse Retrievers learn set-compositional query representations?
    
    \item\label{rq:negation-lsr} How can we improve the representation power of Learned Sparse Retrievers (LSR) to account for negations?
    
\end{enumerate}

\subsection{Zero-shot compositional query retrieval with Linear Algebra Operations}\label{sec:results-rq1}
In this section we try to answer \ref{rq:query-representation}, that aims to build zero-shot compositional representations using atomic queries and their logical operators. We apply a framework that combines atomic LSR embeddings using Linear Algebra Operations (Section \ref{sec:method:rq1}).

\begin{table}
\begin{adjustbox}{width=\columnwidth, center}

\begin{tabular}{ll|ll}
\toprule
method & $\vec{Q_{A \star B}}$ & NDCG@10  & R@100\\\hline
\multicolumn{4}{c}{\textbf{Set Difference $(q_A \setminus q_B)$}} \\\hline
\multirow{2}{*}{vanilla} &  $f_{QE}$($q_A$ but & \multirow{2}{*}{0.132}  & \multirow{2}{*}{0.307}\\
& not $q_B)$
\\\hdashline
score-fusion $(-)$ &  -- & 0.228 & 0.332\\
score-fusion-scaled $(-)$ & -- & 0.228 & 0.332 \\\hdashline

Subtraction &$\vec{A}-\vec{B}$
& 0.120 & 0.190 \\
Ignore negation & $\vec{A}$
& 0.233 &  \textbf{0.390} \\

Orthogonal  & \multirow{2}{*}{$\vec{A} - \pi_{\vec{B}} \vec{A}$ } & \multirow{2}{*}{0.235} & \multirow{2}{*}{0.379} \\
negation~\cite{widdows2003orthogonal}\\

NRF~\cite{salton1990improving} & $\vec{A}-\lambda \cdot \vec{B}$ & \textbf{0.256} & 0.379 \\

Disentangled  & \multirow{2}{*}{$\vec{A}-\vec{B^*}$} & \multirow{2}{*}{\textbf{0.258}} & \multirow{2}{*}{\textbf{0.398}} \\
negation \textbf{(ours)} & \\
\hline

\multicolumn{4}{c}{\textbf{Union $(q_A \cup q_B)$}} \\\hline
vanilla & $f_{QE}(q_A$ or $q_B$)& {0.243} & {\textbf{0.383}} \\\hdashline
score-fusion $(+)$ & -- & 0.204 & 0.335 \\
score-fusion-scaled $(+)$ & -- & 0.197 & 0.326 \\
\hdashline
Addition &$\vec{A}+\vec{B}$  & 0.237 & \textbf{0.386}  \\
Max Pool & $MaxPool(\vec{A},\vec{B})$  & \textbf{0.254} & \textbf{0.387} \\
\hline

\multicolumn{4}{c}{\textbf{Intersection $(q_A \cap q_B)$}} \\\hline
\multirow{2}{*}{vanilla} & $f_{QE}(q_A$ that & \multirow{2}{*}{0.063} & \multirow{2}{*}{0.209}\\
&  are also $q_B)$ 
\\\hdashline

score-fusion $(\cdot)$ & -- & 0.070 & 0.111\\
score-fusion-scaled  $(\cdot)$ & -- & 0.070 & 0.111 \\\hdashline
Addition &$\vec{A}+\vec{B}$  &  0.068 & 0.222 \\
Max Pool &$MaxPool(\vec{A},\vec{B})$ &  0.069 & 0.238 \\

CPT \textbf{(ours)} & $\vec{A} \otimes \vec{B}$ 
& \textbf{0.080} & \textbf{0.268} \\
\hline

\end{tabular}
\end{adjustbox}

\caption{Deriving set-compositional LSR representations with Linear Algebra Operations (LAO) in a zero-shot setting. 
}\vspace{-1em}
\label{tab:results-lao}
\end{table}

\begin{table}
    \centering
    \begin{adjustbox}{width=\columnwidth, center}
    \begin{tabular}{p{1.25cm}p{7cm}}
    method & representation term scores\\\midrule

    \multicolumn{2}{c}{\textit{(Books about monarchs) $\setminus$ (Books about European monarchs)}}\\\midrule
    vanilla & (monarch\#\#: 1.11, \textbf{not: 1.1}, book\#\#: 1.03, \textbf{european: 1.02}, king: 0.38, non: 0.35, ...) \\\hdashline
    
    Negation Ignore &  (monarch\#\#: 1.25, history: 1.13, book\#\#: 1.03, king: 0.77, historical: 0.72, royal: 0.63, ... \st{european: 0.0}) \\\hdashline
    
    Orthogonal negation & (history: 1.12, monarch\#\#: 1.07, book\#\#: 0.89, king: 0.74, royal: 0.61, ..., \st{european: 0.0})\\\hdashline
    
    NRF & ({history: 0.81}, { monarch\#\#: 0.64}, book\#\#: 0.49, king: 0.43,royal: 0.34, ..., \textbf{europe: -0.16}, \textbf{european: -0.29}) 
    \\\hdashline

    Disenta- ngled Negation & ({monarch\#\#: 1.25}, {history: 1.13}, book\#\#: 1.02, king: 0.77,  ..., \textbf{europe: -0.66}, \textbf{european: -1.17}
    \\\hdashline\hdashline
    finetuned (RQ2) & (monarch\#\#: 1.2, king\#\#: 1.04, books: 0.78, tsar: 0.75, encyclopedia: 0.72, book: 0.71, biography: 0.67,
    emperor: 0.65, ..., \st{european: 0.0})  \\

    \midrule
    \multicolumn{2}{c}{\textit{(Documentary films about education) $\cap$}}\\
    \multicolumn{2}{c}{\textit{(Documentary films about intellectual disability) }}\\
    \midrule
    vanilla & (intellectual: 1.14, disability: 1.07, documentary: 1.04, education: 1.04, film\#\#: 0.9,disabled: 0.82, educational: 0.73, ...) \\\hdashline

    CPT & (education $\cap$ intellectual: 1.57, documentary $\cap$ intellectual: 1.49, education $\cap$ disability, 1.47, documentary $\cap$ disability: 1.40, %
    ...
    ) %
    \\\hdashline\hdashline
    finetuned (RQ2) &documentary: 1.21, education: 1.12, disability: 1.08, intellectual: 1.02, disabled: 1.02, school: 0.77, academic: 0.73, autism: 0.67, %
    ...\\
    \midrule

    \end{tabular}
    \end{adjustbox}
    
    \caption{Examples of set-compositional LSR representations. %
    }\vspace{-3em}
    \label{tab:query-analysis}
\end{table}

In Table \ref{tab:results-lao} we measure ranking performance of various zero-shot methods, designed to tackle set-compositionality. Those include the $vanilla$ baseline, score-fusion methods that fuse document scores from queries $q_A$ and $q_B$, and the compositional representations derived using the framework and Linear Algebra Operations (LAO) we introduced in Section \ref{sec:method:rq1}.
Our first observation is that pre-trained LSR models ($vanilla$) work well for Union queries, but can be improved for Set Difference and Intersection queries. 
Score-fusion methods are effective only on Set Difference queries, but do not outperform the simple method of ignoring the negation. 
From the last group of methods, we conclude that constructing compositional representations with the LAO can help for the more challenging operations of Set Difference and Intersection. 

\textit{Set Difference ($A\setminus B$)}.
For Set Difference, subtracting the LSR vectors is ineffective, but ignoring the negation works particularly well, especially in $R@100$. We hypothesize performance of this method is exaggerated due to (a) the small document collection ($\sim300K$ docs) and (b) the synthetic nature of Quest queries (minimal overlap between positive and negative query, discussed in Subsection \ref{sec:results-rq1-negation-proximity}). 
Orthogonal negation performs on par with Negation Ignore, as it subtracts from $\vec{A}$ the projection of $\vec{B}$ in $\vec{A}$. 
This does not change the direction of vector $\vec{A}$, and in practice only slightly affects its' norm $|\vec{A}|$, leading to similar results. Table \ref{tab:query-analysis} shows example representations demonstrating that in practice, term scores between the two change only slightly.
NRF and Disentangled Negation are the most successful methods, with Disentangled Negation having a higher $Recall$ compared to NRF. 
Our hypothesis is that the advantage of these methods in top ranking positions (ie. $NDCG$) stems from their ability to penalize the appearance of negated terms in documents.
However, as seen in  Table \ref{tab:query-analysis}, Disentangled Negation does so in a more targeted way by penalizing the negated terms  more (eg. europe) without lowering the term importance of the positive ones (eg. monarch, history) leading to higher $R@100$. 
We investigate this property further in Section \ref{sec:results-rq1-negation-proximity}.

\textit{Union ($A\cup B$)}.
Similarly to previous work, we find that union queries are less challenging~\cite{malaviya2023quest}. This is intuitive, since by nature Union operations are the most easy to satisfy, ie. by relevance to any of the atomic queries. 
Max Pooling performs slightly better compared to the vanilla variant, with simple Addition being somewhat less effective. 
This is not surprising, given that Splade models already apply a max pooling operation to aggregate across input tokens.

\textit{Intersection ($A\cap B$)}.
Lastly, we find that Intersection is the most challenging operation for LSR models. 
As explained in Section \ref{sec:method:rq1:lao}, traditional lexical methods are unable to capture the combined occurrence of terms across queries that this set operation necessitates. 
These methods can attribute high scores in all the important terms (eg. education, disability, documentary in Table \ref{tab:query-analysis}) but fail to capture their combined existence, which would further help them determine relevance.
Our Combined Pseudo-Term (CPT) method partially addresses this issue by expanding the vocabulary space to consider two term combinations, outperforming all other methods. 

\subsubsection{Effectiveness of negation methods for varying levels of interference between positive and negative queries}\label{sec:results-rq1-negation-proximity}
Previous work has discussed the challenges of applying negations in IR, focusing on the difficulty of removing negative aspects from a query vector without compromising the positive ones~\cite{kowalski2007information,dunlop1997effect}. \citet{widdows2003orthogonal} showed that as the similarity between positive and negative query vectors increases, it becomes harder to account for negations in ranking.
We investigate this further using QUEST, which includes queries with varying levels of interference (e.g., "Books about French monarchs but not Napoleon" vs. "1988 fiction books that are not about the environment").

\begin{figure}[]
    \centering
        \centering
        \includegraphics[width=\columnwidth]{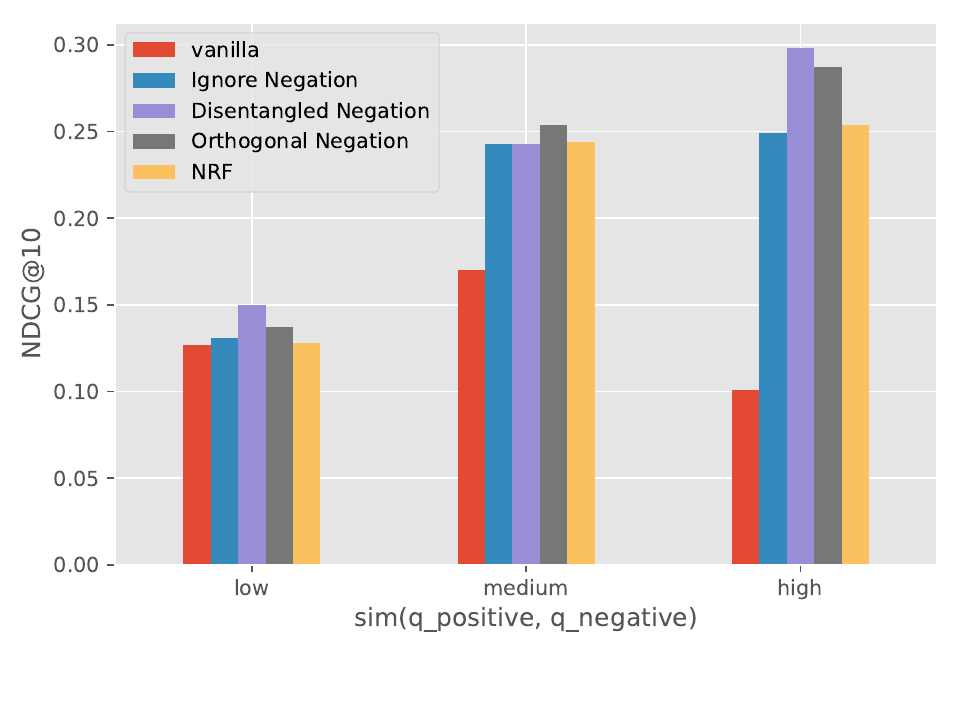}\vspace{-3em}
    \caption{Performance of zero-shot negation methods, at different levels of interference between positive and negative query.}\vspace{-1em}
    \label{fig:negation-similarity}
\end{figure}

We measure method performance in Figure \ref{fig:negation-similarity}, for varying levels of interference (ie. similarity\footnote{Measured with a Dense Retriever \textsc(sentence-transformers/all-MiniLM-L6-v2)}) between positive and negative query vectors.
In low similarity regions, all methods perform relatively on par. However, as the overlap increases, negation methods outperform the \textit{vanilla} variant by a large margin. 
At the highest level of interference, ignoring the negation or subtracting a constant value of the negative query still work reasonably well, but Disentangled and Orthogonal Negation outperform them by a clear margin. 
This is due to the ability of these methods to remove the negative aspects of the query without harming the positive aspects, which is a desirable and important property for negations.

\subsubsection{What's the role of expansion and term weighting in the success of linear algebra operations?}

LSR improves ranking over lexical methods by introducing two mechanisms that enhance the matching semantics: (a) lexical expansion, that bridges the vocabulary mismatch between queries and documents, and (b) term scoring, that re-calibrates query-document scores by estimating the term importance of individual tokens rather than relying in global term statistics (eg. BM25~\cite{robertson2009probabilistic}). 
In this part, we want to explore to what extent these two mechanisms affect the effectiveness of the linear algebra operations.
\begin{table*}

\begin{tabular}{l||r|r|r%
}
\toprule

 & \textbf{LSR (full)}\footnote{https://huggingface.co/naver/splade-v3}
&  \textbf{LSR (w/o Q.Exp.)}\footnote{https://huggingface.co/naver/splade-v3-lexical/}
&  \textbf{LSR-doc} \footnote{https://huggingface.co/naver/splade-v3-doc/}
\\

\multicolumn{1}{r||}{Query Exp.} & \ding{52} & \ding{54} &  \ding{54} 
\\

\multicolumn{1}{r||}{Query Term Scoring} &  \ding{52} & \ding{52} & \ding{54} 
\\

\multicolumn{1}{r||}{Doc. Exp.}&\ding{52} & \ding{52} & \ding{52} 
\\

\multicolumn{1}{r||}{Doc. Term Scoring} &\ding{52} & \ding{52} & \ding{52} 
\\
\toprule\toprule

\textbf{Query type} &  \multicolumn{3}{c}{$NDCG@10$ 
    \hspace{3.5em} } \\ 
\multicolumn{1}{r||}{+Applied Operation} &  \multicolumn{3}{c}{
    \hspace{3.5em}$(\pm \Delta NDCG \%)$} \\\hline

\textbf{Atomic} & \multicolumn{1}{l|}{0.284} & \multicolumn{1}{l|}{0.264} & \multicolumn{1}{l|}{0.258} 
\\\hdashline

\textbf{Set Difference} & \multicolumn{1}{l|}{0.155} & \multicolumn{1}{l|}{0.125} & \multicolumn{1}{l|}{0.073} 
\\
\multicolumn{1}{r||}{\hspace{1em}+Disentangled Negation} & 0.258 (\textbf{+66.5\%}) & 0.206 (\textbf{+64.8\%})& 0.210 (\textbf{+187\%})\\\hdashline

\textbf{Union} & \multicolumn{1}{l|}{0.250} & \multicolumn{1}{l|}{0.236} & \multicolumn{1}{l|}{0.231} 
\\
\multicolumn{1}{r||}{\hspace{1em}+MaxPool} & 0.262 (\textbf{+4.8\%}) & 0.251 (\textbf{+6.4\%}) & 0.233 (+0.9\%) \\\hdashline

\textbf{Intersection} & \multicolumn{1}{l|}{0.072} & \multicolumn{1}{l|}{0.065} & \multicolumn{1}{l|}{0.069} 
\\
\multicolumn{1}{r||}{\hspace{1em}+CPT}& 0.098 (\textbf{+36.1\%}) & 0.087 (\textbf{+33.8\%}) & 0.065 (-5.8\%) \\
\bottomrule

\end{tabular}

\caption{
Investigating the effect of (a) Query Expansion, (b) Query Term Scoring, (c) Document Expansion and Document Term Scoring
in the effectiveness of LAO.
Absolute numbers (first row of group, aligned left) indicate vanilla $NDCG@10$ performance for each model, while $\pm\Delta \%$ (aligned right) indicate relative differences after applying the linear algebra operations.
}
\label{tab:ablation-lsr-bm25}
\end{table*}

On Table \ref{tab:ablation-lsr-bm25} we incrementally remove (a) query expansion, (b) query term scoring, and (c) document expansion and term scoring from LSR models
\footnote{Here, we use the splade-v3~\cite{lassance2024splade} models, that are available for different expansion and term scoring settings}
. 
By removing those components incrementally in the vanilla variant, we observe that performance degrades slightly, following the trend seen in atomic queries. 
This is not the case on Set Difference queries, where a large performance drop occurs when removing Query Term Scoring ($0.125$ vs. $0.073$), in practice using a binary query vector. 
This suggests that Splade has (to some extent) learned how to handle negations during their pre-training (MSMarco or BERT pretraining), lowering the scores of negated terms or completely removing them.

Steering our focus towards the improvements of LAO in models with and without expansion and term scoring, 
we observe that removing Query Expansion does not substantially affect those improvements.
However, those improvements collapse when we remove query term scoring (LSR-doc), with the exception of negated queries. 
For CPT this is expected, as atomic term scores are propagated towards the final query representation through the outer product (Eq. \ref{eq:intersection-outer-product}), ranking higher documents that contain the CPT (\textsc{Venezuela}$ \cap $\textsc{Colombia}), instead of (\textsc{Venezuela}$ \cap $\textsc{Andes}).

In this section, we introduced an effective zero-shot framework that constructs set-compositional representations using Linear Algebra Operations. 
The proposed operations of Disentangled Negation and Combined Pseudo-Terms outperform the vanilla by a large margin, indicating the effectiveness of our framework.

\subsection{Can retrievers learn query compositional representations?}\label{sec:results-rq2}

\begin{table*}[]
\begin{tabular}{l|cc|cc|cc|cc}
\toprule
train\_set &  \multicolumn{2}{c}{Atomic} &
\multicolumn{2}{c}{Set Difference} &\multicolumn{2}{c}{Union} & \multicolumn{2}{c}{Intersection}\\
 &  $NDCG@3$ & $R@100$  
 &  $NDCG@3$ & $R@100$  
& $NDCG@3$ & $R@100$ 
& $NDCG@3$ & $R@100$ \\\hline

& \multicolumn{8}{c}{Lexical Retrieval} \\\hline
-- & 0.216 & 0.456 & 0.052 & 0.146 & 0.198 & 0.368 & 0.047 & 0.200 \\

\hline & \multicolumn{8}{c}{Learned Sparse Retrieval} \\\hline
Quest-NC $^{a}$ & 

0.219 & 0.458 & 
0.132 & 0.307  &
0.243 & 0.383 & 
0.063 & 0.209 
\\ %
Quest-FULL $^{b}$ & 

\textbf{0.243}$^{a}$ & \textbf{0.499}$^{a}$ & 
0.240$^{a}$ & \textbf{0.409}$^{a}$ &
\textbf{0.260} & \textbf{0.404}$^{a,c}$ &
0.062 & 0.227
\\\hdashline

 \begin{tabular}{@{}l@{}}
   Quest-NC + LAO $^{c}$\\
   (best of RQ1)\\
 \end{tabular} & -- & -- & \textbf{0.258}$^{a}$ & 0.398$^{a}$ & 0.254 & 0.387 & \textbf{0.080}$^{a,b}$ & \textbf{0.268}$^{a,b}$ \\

\hline
& \multicolumn{8}{c}{Dense Retrieval} \\\hline
Quest-NC $^{d}$& 

0.289 & 0.571 & 
0.154 &  0.356 & 
0.278 &  0.446 & 
\textbf{0.088} &  \textbf{0.290} 
\\

Quest-FULL $^{e}$ & 
\textbf{0.311}$^{d}$ & \textbf{0.577} & 
\textbf{0.251}$^{d}$ & \textbf{0.450}$^{d}$ &
\textbf{0.291}$^{d}$ & \textbf{0.466}$^{d}$ & 
0.081 & 0.283
\\
\bottomrule
\end{tabular}
\caption{To what extent can different retriever families learn compositional representations from training? (RQ2). 
Superscripts indicate significant improvements, compared to retrievers of the same family (p-value<0.05, paired t-test).
}\vspace{-1em}
\label{tab:training-compositionality}
\end{table*}

After investigating zero-shot representations, we explore whether effective representations can be learned by Dense and Learned Sparse Retrievers from training%
. 
To answer \ref{rq:learn-representations}, we finetune retrievers on (a) Quest-NC, the non-compositional subset of Quest containing atomic queries and (b) Quest-FULL, that also contains set-compositional queries and present results in Table \ref{tab:training-compositionality}. In general, we observe that Dense Retrievers seem to have a better performance in Quest, both when it comes to atomic or compositional queries.

\textit{Retrievers can effectively learn to handle negated queries, but not intersections.}
Adding set-compositional queries to the training set seems to improve performance on atomic queries by roughly $10\%$, likely due to a larger training set.
Large performance gains are observed only for set difference queries, where LSR learns to remove negated terms from query representations (Table \ref{tab:query-analysis}).
Improvements for union are significant, but comparable to the ones observed for atomic. 
Interestingly, none of the retrievers seem to learn from training on intersections. This aligns with prior research showing that Dense Retrievers struggle with intersections~\cite{malaviya2023quest,dasgupta2021word2box}, but further reveals that, under some domain shift (same dataset, different entity type), both Dense and LSR models fail to effectively model intersections.
Although this is expected for LSR due to their limited representation power in creating term combinations (Section \ref{sec:method:rq1:lao}), it is a surprising finding for DR models that can create high-dimensional representations without sparsity  constraints or lexical grounding.
To further solidify these findings, we also experimented using BM25 and compositional negatives,%
but found no substantial difference (Appendix \ref{sec:appendix:hard-negatives}, Table \ref{tab:training-hard-negatives}). 

\textit{Zero-shot vs. finetuned representations}
Finally, we observe that applying LAO between LSR embeddings achieves performance comparable or better than training them for compositionality.
indicating that their representation power may be in some ways limited. 
For intersections, solving this representation bottleneck would require expanding the vocabulary (ie. Combined Pseudo Terms), or abandoning the semantic relationship between sparse dimensions and tokens leading to a Dense Retriever with more dimensions.
Similarly, we observe that the zero-shot \textit{Disentangled Negation} outperforms both trained LSR and DR models on Set Difference queries in NDCG and remains competitive in Recall. 
Our hypothesis is that negative term weights benefit ranking, as they allow the scoring function to go beyond ignoring negated terms (eg. ``european'', Table \ref{tab:query-analysis}) and penalize documents that contain them.

\subsection{Improving LSR representations for negations}\label{sec:results-rq3}

\begin{table}[]
    \centering
    \begin{tabular}{cccc|c}
      &       &   Negative  &   +/-  &    \\
      & FT on & term & weight &  Pairwise \\
    Model & NevIR & scores & aggreg. &  Accuracy\\
    \midrule
    
    random & \xmark& -- & --&25.00\% \\
    \midrule
    
    \multirow{2}{*}{{contriever}} & \xmark
    & -- & --& 9.62\% \\
    & \cmark & -- & --&42.88\% \\
    \midrule
    \multirow{4}{*}{splade} & \xmark 
    & \xmark & -- & 7.88\% \\ 
    & \cmark & \xmark & -- & 23.07\%  \\
    
    & \cmark& \cmark & $\pm|max|$ & 37.09\% \\ %
    
    & \cmark& \cmark & $sum$ & \textbf{42.73\%} \\
    \end{tabular}
    
    \caption{Performance of LSR models trained with negative term scores on NevIR. %
    }\vspace{-2em}
    \label{tab:rq3-nevir}
\end{table}

Previous experiments %
indicated that LSR representations that attribute \textit{negative} term scores to negated aspects of a query (ie. Disentangled Negation, NRF) seem to have an advantage over methods that simply ignore the negation. Further LSR models seem well-suited for addressing the problem of interference between positive and negative queries.
Therefore, we aim to answer \ref{rq:negation-lsr},
that is, whether the representation power of LSR models can be improved for negations by allowing them to attribute \textit{negative term scores}.

In Table \ref{tab:rq3-nevir}, we measure performance on NevIR~\cite{weller2023nevir} and conclude that even after finetuning, LSR models without negative term weights cannot even surpass the random baseline ($23.07\%$ vs $25\%$). 
In contrast, allowing splade models to score terms negatively helps them almost double their performance on NevIR ($42.73\%$ vs. $23.07\%$). 
Aggregating the maximum positive and negative query scores by adding them ($sum$ weight aggregation), leads to better performance compared to selecting the absolute maximum ($\pm |max|$ weight aggregation). 
This is possibly due to the fact that, in this variant, gradients flow back from both the positive and negative part of the query to the $out$ embedding, leading to a more informative training procedure.

\begin{table}[]
    \centering
    \begin{tabular}{cc|ccccc}
Nega-  &    +/-  &   & \\
tive  & weight   & \multicolumn{2}{c}{Atomic} & \multicolumn{2}{c}{Set Difference} \\
scores & aggreg. & NDCG@10 & R@100 & NDCG@10 & R@100 \\\hline
    
  \xmark & --             & \textbf{0.284} & \textbf{0.540} & 0.205 & \textbf{0.413} \\
  \cmark & $\pm|max|$     & 0.182 & 0.418 & 0.210 & 0.340 \\
  \cmark & sum            & 0.165 & 0.391 & \textbf{0.218} & 0.351 \\
    
    \end{tabular}
    
    \caption{Performance of LSR models trained with negative term scores on Quest. %
    }\label{tab:rq3-quest}
    \vspace{-2em}
\end{table}

Table \ref{tab:rq3-quest} reports ranking performance on the Quest dataset. 
We observe that models with negative weights struggle on atomic queries and do not perform on par with the baseline that uses only positive weights. 
On Set Difference queries, allowing negative term scores improves performance for NDCG but degrades in Recall. %

Our hypothesis for the observed degradation is that by changing the activation function of splade (Figure \ref{fig:activation-functions}, Equations \ref{eqn:splade-activation-vanilla},  \ref{eq:SNRelu-maxV2} and \ref{eq:SNRelu-maxV3}) we introduce a drastic change in the architecture, 
that is hard to learn with a relatively small training set ($1896$ negative train queries on NevIR and $236$ on Quest).
That is due to pre-training of LSR models in the Masked Language Modelling task (BERT~\cite{devlin2018bert}), as well as their pre-finetuning in ranking, where models
are trained to predict very low values on the MLM output $out_{ij}$ 
(left region of Figure \ref{fig:activation-functions}) for irrelevant terms,
and large positive values (right region, Figure \ref{fig:activation-functions}) for both positive and negative terms (see ``non-European'' for \textit{vanilla} variant, Table \ref{tab:query-analysis}). 
Therefore, when initializing model weights with the new activation function (\textsc{SNReLU}), there needs to be a re-calibration of the learned notions of \textit{relevance}, \textit{negative relevance} and \textit{irrelevance}. 
This adjustment requires shifting negatively relevant terms from the positive region of Figure \ref{fig:activation-functions} to the negative (left region), and irrelevant terms, that hold large negative values to the region around $0$ ($out_{ij} \in [-\epsilon,+\epsilon]$) so that the representation remains sparse.
It seems that such an adaptation can be to some extent learned in NevIR, where a relatively larger training set with hard contrastive negatives exists, but is challenging to achieve in Quest. 

To conclude, we find that the representation power of LSR models can be improved by allowing models penalize negatively relevant terms, given a large enough dataset to finetune model weights.

\section{Conclusions and future directions}

In this work, we focused on constructing set-compositional and negated representations for first stage ranking, primarily focusing on LSR representations that are lexically grounded and easier to manipulate. 
We found that constructing effective representations is possible using a zero-shot framework that combines atomic representations using Linear Algebra Operations %
. 
We introduced LAO that target specific set operations, more importantly: \textit{Disentangled Negation} for Set-Difference, that aims to tackle the challenging problem of interference between positive and negative queries,
and Combined Pseudo-Terms that aims to expand the representation space of Intersections by considering term combinations. We found both to be particularly effective, performing on par or better compared to models trained for set-compositionality.
When it comes to training, we found that both LSR and DR cannot learn effective representations for intersections, but do benefit from training on queries with negations. 
Finally, we adapted LSR models for attributing negative term scores to negated aspects, that in practice allows them to effectively penalize documents containing those aspects. 
We find this to be particularly effective in settings where a reasonably sized negation dataset exists, but more challenging to apply when data is scarce. 

This work lays the foundation for creating lexically grounded, set-compositional and negated LSR representations. 
We believe that numerous exciting future directions  exist, with the most important being: 
(a) researching ways on making LSR models with negative term scores more effective under data scarce settings, possibly by pre-training methods and data generation using LLMs, or knowledge distillation from existing retrievers and rerankers,
(b) exploring the efficacy of late-interaction models in creating set-compositional representations, due to their more expressive representations (ie. one Dense vector per query token), 
(c) investigating further the problem of query decomposition with LLMs and exploring whether a more efficient query encoder can also perform this task with high accuracy.

\bibliographystyle{ACM-Reference-Format}
\balance
\bibliography{main}

\clearpage

\appendix

\section{Can retrievers benefit from hard negative mining to learn compositional representations}\label{sec:appendix:hard-negatives}

\begin{table*}
\begin{tabular}{llcccccc}
\toprule
train\_set & negatives &
\multicolumn{2}{c}{Set Difference} &\multicolumn{2}{c}{Union} & \multicolumn{2}{c}{Intersection}\\
& &  $NDCG@3$ & $R@100$  
 &  $NDCG@3$ & $R@100$  
& $NDCG@3$ & $R@100$ \\\hline

\multicolumn{8}{c}{Learned Sparse Retrieval} \\\hline
\multirow{2}{*}{Quest-NC} & inbatch & 
0.132 & 0.307 &
0.243 & 0.383 & 
0.063 & 0.209 \\ 
 & \hspace{1em} + BM25 & 
\textbf{0.156} & \textbf{0.359} & 
\textbf{0.278} & \textbf{0.436} &  
\textbf{0.069} & \textbf{0.250} \\\hdashline

\multirow{2}{*}{Quest-FULL} & inbatch & 
\textbf{0.240} & \textbf{0.409} &
\textbf{0.260} & \textbf{0.404}&
\textbf{0.062} & \textbf{0.227} \\

 & \hspace{1em} + BM25 & 
0.232 & 0.399 &
0.237 & 0.387 &
0.058 & 0.207 \\ %

\hline\multicolumn{8}{c}{Dense Retrieval} \\\hline
\multirow{2}{*}{Quest-NC} & inbatch & 
\textbf{0.154} &  \textbf{0.356} &
\textbf{0.278} &  \textbf{0.446} & 
\textbf{0.088} &  \textbf{0.290} \\

 & \hspace{1em} + BM25 & 
0.150 &  0.341 &
0.257 &  0.432 &    
0.074 &  0.272 
 \\\hdashline

\multirow{4}{*}{Quest-FULL} & inbatch & 
\textbf{0.251} & \textbf{0.450} &
\textbf{0.291} & \textbf{0.466} & 
\textbf{0.081} & \textbf{0.283} 
\\

 & \hspace{1em} + BM25 & 
0.246 &  0.428 &
0.278 &  0.428 &      
0.074 &  0.269 \\
 & \hspace{1em} + composit. & 
0.249 &  \textbf{0.455} &
0.289 &  0.456 &   
\textbf{0.081} &  0.285 \\
 & \hspace{1em} + both & 
0.249  &  0.422  &
0.277  &  0.436  &    
0.080  &  0.266 \\
\bottomrule
\end{tabular}
\caption{Can hard negatives help retrievers learn better compositional representations? (RQ2)}
\label{tab:training-hard-negatives}
\end{table*}

Hard negative mining has been one of the most important areas of focus for improving Dense and Learned Sparse Representations~\cite{formal2021spladev2,yates2021pretrained}. To this end, we explore whether hard negative mining can improve compositional representations build by retrievers. 
We experiment with BM25 negatives, as well as compositional negatives.
We sample BM25 negatives randomly from ranking positions $50-100$.
We also construct hard compositional negatives, ie. documents that are related to one of the atomic queries, but not the final compositional query.
Specifically, those are: 
\begin{itemize}
    \item For Set Difference ($q_{A \setminus B}$), documents relevant to $q_B$ ($|B|$)
    \item For Intersection ($q_{A \cap B}$), documents relevant to one of the atomic queries, but not both for Intersection$|A \cup B| \setminus |A \cap B|$ 
\end{itemize}

We measure performance on Table \ref{tab:training-hard-negatives}, but conclude that these negatives do not seem to improve ranking performance consistently for any compositionality type. 
This might be attributed to the limited representation power of rankers for certain compositionality types (eg. LSR for intersection), the limited amount of training data, or the training techniques used (eg. ranking loss).

\end{document}